\documentstyle[a4,12pt,amssymb]{article}
\input epsf
\begin{document}
\baselineskip=20pt
{\center
{\bf \huge Improving a method for 
the study of limit cycles of the Li\'enard equation\\}
\vspace{1 cm}
{\large Hector Giacomini
\footnote{email: giacomini@univ-tours.fr}
 and S\'ebastien Neukirch 
\footnote{email: seb@celfi.phys.univ-tours.fr}\\
Laboratoire de Math\'ematiques
 et Physique Th\'eorique\\ C.N.R.S. UPRES A6083 \\
Facult\'e des Sciences et Techniques, Universit\'e de Tours\\
F-37200 Tours FRANCE\\
}}
\vspace{3 cm}
{\center \section*{Abstract}}
{ \small
In recent papers we have introduced a method
for the study of limit cycles of the Li\'enard
system~: $\dot{x}=y-F(x) \; , \; \dot{y}=-x$ , where $F(x)$
is an odd polynomial. The method gives a sequence
of polynomials $R_n(x)$, whose roots are related
to the number and location of the limit cycles,
and a sequence of algebraic approximations to the
bifurcation set of the system.
In this paper, we present a variant of the
method that gives very important qualitative and 
quantitative improvements.}

\vspace{5 cm}

{\bf PACS numbers : 05.45.+b , 02.30.Hq , 02.60.Lj , 03.20.+i 

Key words :  Li\'enard equation, limit cycles}
\clearpage
In a previous paper \cite{h et s}, we have introduced
a method for studying the number and location of
limit cycles of the Li\'enard system~:
\begin{equation}
\frac{d x}{d t}=y-F(x) \quad , \quad \frac{d y}{d t}=-x \quad , 
\label{eq lienard}
\end{equation}
where $F(x)$ is an odd polynomial of arbitrary degree.
The method is as follows~: we consider a function $h_n(x,y)$
given by~:
\begin{equation}
h_n(x,y)=y^n+g_{n-1,n}(x) y^{n-1}+g_{n-2,n}(x) y^{n-2}
+...+g_{1,n}(x) y + g_{0,n}(x) \quad , \nonumber
\end{equation}
where $g_{j,n}(x)$, with $j=0,1,...,n-1$, are functions of $x$ only
and $n$ is an even integer.
It is always possible to choose the functions $g_{j,n}(x)$
such that~:
\begin{equation}
\frac{d}{dt}h_n(x,y) \equiv \dot{h}_n(x,y)=(y-F(x))
\frac{\partial h_n}{\partial x} -x \frac{\partial h_n}{\partial y}
\end{equation}
is a function of the variable $x$ only (see also \cite{cherkas}).
Hence we have~:
\begin{equation}
\dot{h}_n(x,y)=R_n(x) \nonumber
\end{equation}

The functions $g_{j,n}(x)$ and $R_n(x)$ determined in this way
are polynomials.
We have shown in \cite{h et s} and \cite{h et s2} that the
polynomials $h_n(x,y)$ and $R_n(x)$ give a lot of information
about the number and location of the limit cycles of (\ref{eq lienard}).
In particular, we have established in \cite{h et s} the following
conjecture~:\\
{\bf Conjecture}~: Let $L$ be the number of limit cycles
of (\ref{eq lienard}). Let $r_n$ be the number of
positive roots of $R_n$ (with $n$ even) of odd multiplicity.
Then we have~:
\begin{itemize}
\item $ L \leq  r_n \nonumber $
\item if $ m > n$ then $ r_m-r_n=2 p$ with $ p \in N $
\end{itemize}
Moreover, we have also shown in \cite{h et s} and \cite{h et s2}
that the polynomials $h_n(x,y)$ and $R_n(x)$ allow us to
construct algebraic approximations to each limit cycle and to
the bifurcation curves of (\ref{eq lienard}).
For the bifurcation set, these algebraic approximations
are exact lower bounds and seem to converge in a monotonous
way to the exact bifurcation set of the system.
The fundamental aspect of this method is that it is not
perturbative in nature. It is not necessary to have a small or
a large parameter in order to apply it.\\

In the present paper, we want to improve the results presented
in \cite{h et s} and \cite{h et s2}.
Let us consider, as a first example, the van der Pol system~:
\begin{eqnarray}
\dot{x} & = & y-\epsilon(x^3/3-x) \nonumber \\
\dot{y} & = & -x \label{eq vdp}
\end{eqnarray}
In this case we have $F(x)=\epsilon (x^3/3-x)$
and the polynomials $R_n(x,\epsilon)$ have only one
positive root of odd multiplicity for each even value
of $n$ and for arbitrary values of the parameter $\epsilon$
(we have indicated here the explicit dependence of $R_n$ in
$\epsilon$ by writing $R_n(x,\epsilon)$).
We call {\it amplitude} of the limit cycle, the maximum value
of the coordinate $x$ on the limit cycle and we will 
refer to it by $x_{max}$. For the van der Pol equation,
 this amplitude is
a function of $\epsilon$ and we will write~:
 $x_{max}(\epsilon)$

In table (\ref{tbl vdp1}), we give the roots of the polynomial
$R_n(x,\epsilon)$ for $n$ between 2 and 20 and for $\epsilon=3$.
These sequence of roots seems to converge in a monotonous
 way to the
amplitude of the limit cycle $x_{max}(\epsilon) \simeq 2.023$,
 which is
obtained by a numerical integration of the system.

As it is explained in \cite{h et s}, the integral of each
 polynomial
$R_n(x,\epsilon)$ along the limit cycle must be zero for
 all even values of $n$~:
\begin{equation}
\int_{0}^{T} R_n(x(t),\epsilon) dt=0 \quad , \label{eq integrale}
\end{equation}
where $T$ is the period of the limit cycle.

Let us now describe for this example the improved method
that represents the new contribution of this paper. We
employ an idea utilized in the averaging method \cite{verhulst}~:
we replace $x(t)$ by $a \cdot \cos(t)$ in (\ref{eq
 integrale}),
where $a$ is an unknown constant, and we replace
the period $T$ by $2 \pi$. After integration, we obtain a
polynomial in $a$ which we denote 
$\hat{R}_n(a,\epsilon)$~:
\begin{equation}
\hat{R}_n(a,\epsilon)=\int_{0}^{2 \pi} R_n(a \cdot \cos(t),\epsilon) dt
\end{equation}
Surprisingly enough, the polynomials $\hat{R}_n(a,\epsilon)$
have the same qualitative properties as the polynomials
$R_n(x,\epsilon)$. Each of them has only one positive root
of odd multiplicity for arbitrary values of $\epsilon$.
The values of these roots, for $n$ between 2 and 20 are
given in table (\ref{tbl vdp2}) for the case $\epsilon=3$.
We can verify that each one of these roots represents
a lower bound for $x_{max}(\epsilon=3)$. This sequence
of roots seems to converge to $x_{max}(\epsilon=3)$
much more rapidly than the sequence of roots of the
polynomials $R_n(x,\epsilon)$ and they represent excellent
approximations to the value of the amplitude of the limit
 cycle.

This behavior of the roots of the polynomials
$\hat{R}_n(a,\epsilon)$ is merely an {\it experimental} fact.
At present, we have no rigorous arguments to explain
these results.
We have observed this behavior of the roots of the
polynomials $\hat{R}_n(a)$ for other Li\'enard systems of 
type (\ref{eq lienard}) and
the conjecture established in \cite{h et s} (and given
also above) about the roots of the polynomials
$R_n(x)$ seems to be valid also for the roots
of the `` averaged '' polynomials $\hat{R}_n(a)$.

\begin{figure}[p]
$$
\epsfxsize=8cm
\epsfbox{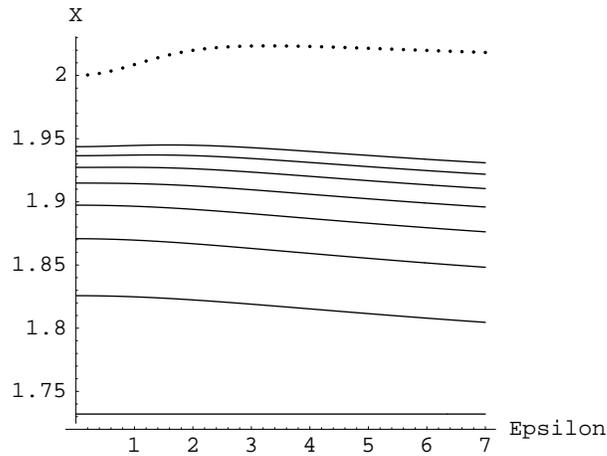}
$$
\caption{Plots of $R_n(x,\epsilon)=0$ for $n=2$ to $n=16$ for
system (\ref{eq vdp}). The point line is the $x_{\max}(\epsilon)$
calculated numericaly.}
\label{fig vdp1}
\end{figure}

\begin{figure}[p]
$$
\epsfxsize=8cm
\epsfbox{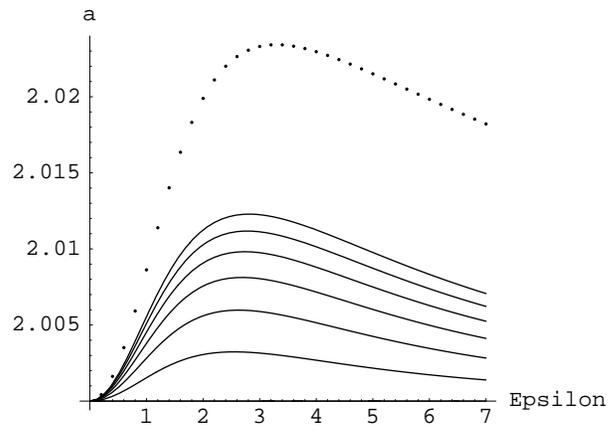}
$$
\caption{Plots of $\hat{R}_n(a,\epsilon)=0$ for $n=2$ to $n=16$
for system (\ref{eq vdp}). The point line is the $x_{\max}(\epsilon)$
calculated numericaly. Note that, here, the horizontal axe is $a=2$.}
\label{fig vdp2}
\end{figure}

For a given value of $n$, we can obtain the 
approximation of the amplitude $x_{max}(\epsilon)$
as a function of $\epsilon$ by considering
the curve given by the
equation $R_n(x,\epsilon)=0$. However, a better approximation
is found by considering $\hat{R}_n(a,\epsilon)=0$
instead.
In fig.(\ref{fig vdp1}) (respectively 
(\ref{fig vdp2})), we give the curve
$R_n(x,\epsilon)=0$ (respectively 
$\hat{R}_n(a,\epsilon)$=0) for several values
of $n$ and the numerical curve $x_{max}(\epsilon)$
obtained from a numerical integration of the system.
As we can see from these figures, the improvement
obtained with the new method is very important
and has two different aspects~:
\begin{itemize}
\item a qualitative aspect~: the curves 
$\hat{R}_n(a,\epsilon)=0$ are nearer to the 
numerical curve than the curve $R_n(x,\epsilon)=0 $.
\item a qualitative aspect~: the asymptotic
behavior (when $\epsilon \to \infty$ or $\epsilon \to 0$)
of the curves $\hat{R}_n(x,\epsilon)=0$ is the correct one
(even for small $n$).
\end{itemize}
The amplitude $x_{max}(\epsilon)$ of the limit
cycle of the van der Pol equation tend to the value 2
when $\epsilon \to \infty$ or $\epsilon \to 0$~:
$$ \lim_{\epsilon \to \infty} x_{max}(\epsilon)=
\lim_{\epsilon \to 0} x_{max}(\epsilon)=2$$
This asymptotic behavior is correctly given
by the curves $\hat{R}_n(a,\epsilon)=0$ for
all even values of $n$.

Despite the fact that the curves $R_n(x,\epsilon)=0$
do not have the correct asymptotic behavior, each one
represent an exact lower bound to the function 
$x_{max}(\epsilon)$ and is closer to
it than its predecessor. Moreover, for a given 
value of $\epsilon$, if we take $n$ sufficiently large
, the root of $R_n(x,\epsilon)=0$ will be as near as we
want to $x_{max}(\epsilon)$.\\
\\
For other recent results about
the limit cycle of the van der Pol equation see \cite{odani}
and \cite{odani2}.\\

Let us consider a second example~:
\begin{eqnarray}
\dot{x} & = & y- \epsilon \, (x^5-\sqrt{\alpha} x^3+x)
 \nonumber \\
\dot{y} & = & -x \label{eq rychkov}
\end{eqnarray}
where  $\epsilon$  and  $\alpha$  are arbitrary positive parameters.
This system has been carefully studied by Rychkov 
\cite{rychkov} and can have at most two limit cycles.
Since there are two parameters, the bifurcation set is
given by a curve in the parameter plane $(\epsilon,\alpha)$.

In \cite{h et s2}, we have shown that the method presented in
\cite{h et s} allows us to obtain a sequence of exact algebraic 
lower bounds to the bifurcation set of systems like 
(\ref{eq rychkov}). Here, we will show that by using the
polynomials $\hat{R}_n(a)$, instead of the polynomials
$R_n(x)$, we can considerably improve the results presented
in \cite{h et s2}.

In the first quadrant of the plane $(\epsilon,\alpha)$, there
exists a bifurcation curve $B(\epsilon,\alpha)=0$. On this 
curve, the system undergoes a saddle-node bifurcation (see
\cite{h et s2} for a description of this type of bifurcation).
Obviously, this function $B(\epsilon,\alpha)=0$ is not known
and no analytical method for obtaining it for arbitrary values
of the parameters exists.

We will obtain algebraic approximations to the curve
$B(\epsilon,\alpha)=0$ from the polynomials 
$R_n(x,\epsilon,\alpha)$ and $\hat{R}_n(a,\epsilon,\alpha)$.
We will call  $B_n(\epsilon,\alpha)=0$  the algebraic
approximations obtained from the polynomials
$R_n(x,\epsilon,\alpha)$
and $\hat{B}_n(\epsilon,\alpha)=0$ the curves obtained from
the polynomials $\hat{R}_n(a,\epsilon,\alpha)$.
As explained in \cite{h et s2}, the function $B_n(\epsilon,\alpha)$
is obtained from the conditions~:
\begin{equation}
R_n(x,\epsilon,\alpha)=0 \; , \; 
\frac{dR_n}{dx}(x,\epsilon,\alpha)=0 \label{eq resu1}
\end{equation}
In the same way, the function $\hat{B}_n(\epsilon,\alpha)$ is
obtained from the conditions~:
\begin{equation}
\hat{R}_n(a,\epsilon,\alpha)=0 \; , 
\; \frac{d\hat{R}_n}{dx}(a,\epsilon,\alpha)=0 \label{eq resu2}
\end{equation}

\begin{figure}[p]
$$
\epsfxsize=8cm
\epsfbox{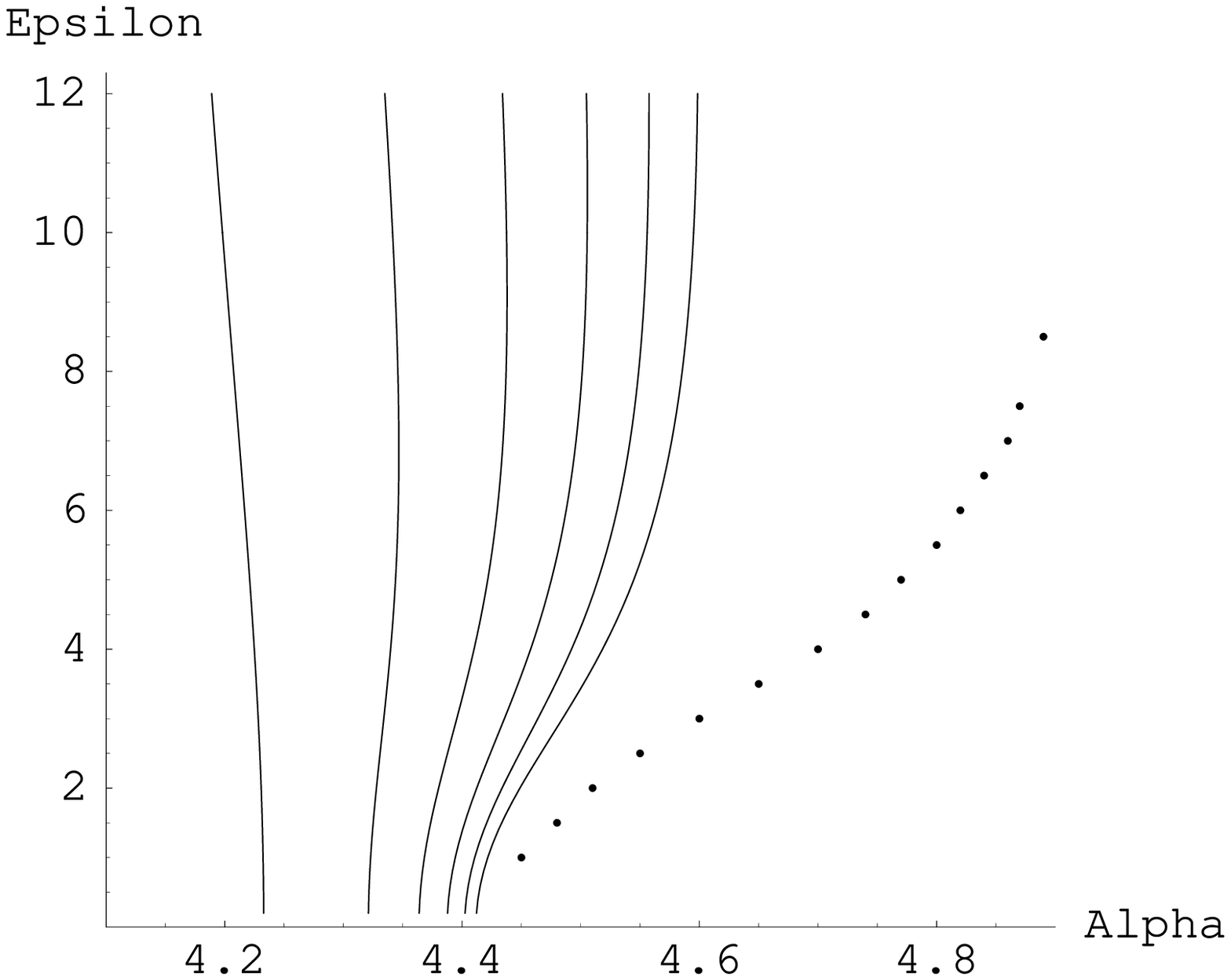}
$$
\caption{Plots of $B_n(\alpha,\epsilon)=0$ for $n=2,6 \mbox{ and } 10$
for system (\ref{eq rychkov}). The point line is 
$B(\alpha,\epsilon)=0$ calculated numericaly.}
\label{fig rychkov}
\end{figure}

\begin{figure}[p]
$$
\epsfxsize=8cm
\epsfbox{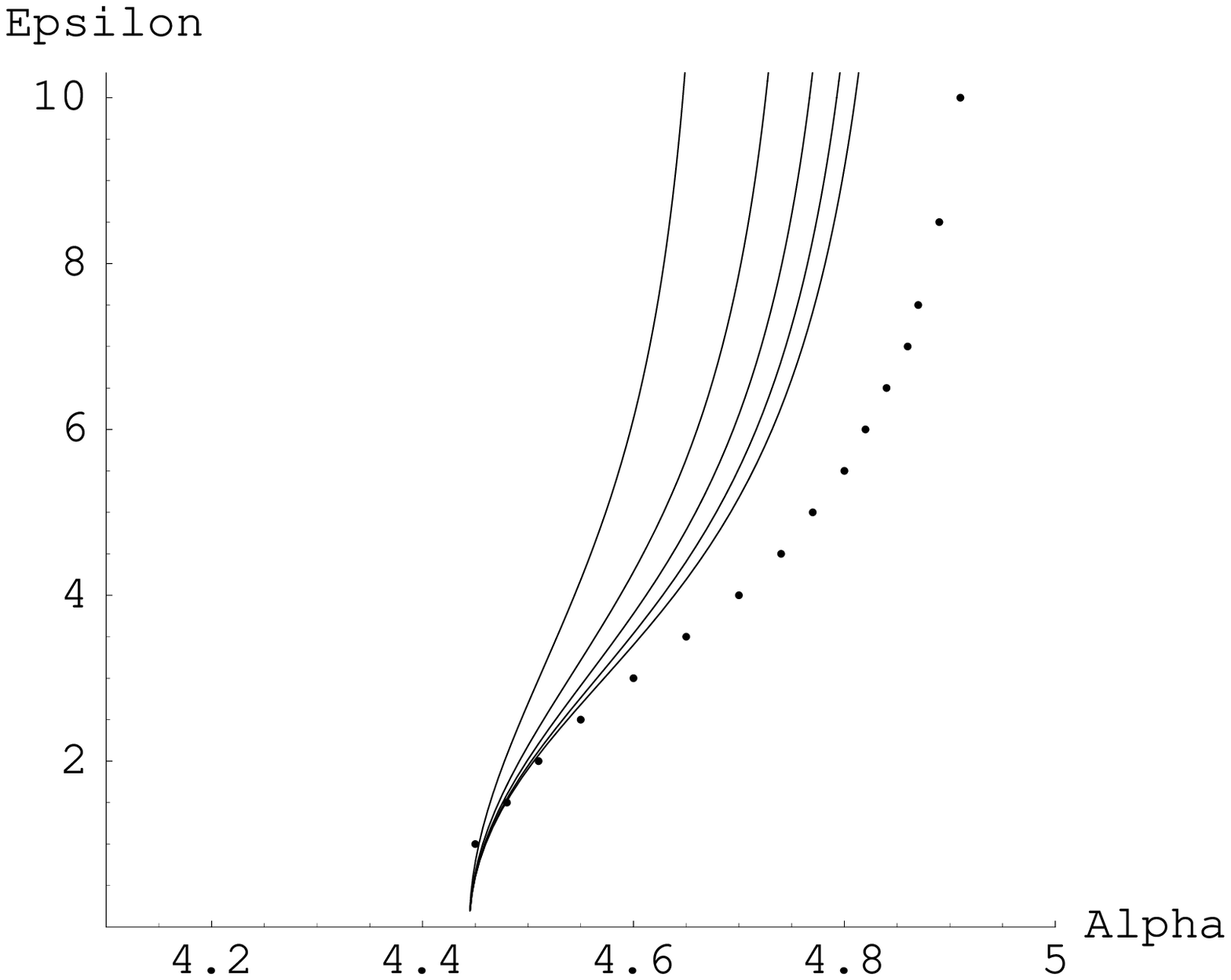}
$$
\caption{Plots of $\hat{B}_n(\alpha,\epsilon)=0$ for $n=2,6 \mbox{ and } 10$
for system (\ref{eq rychkov}). The point line is 
$B(\alpha,\epsilon)=0$ calculated numericaly.}
\label{fig rychkov2}
\end{figure}

The algebraic equations (\ref{eq resu1}) 
(respectively (\ref{eq resu2})) 
determine the double root of $R_n(x,\epsilon,\alpha)$
(respectively $\hat{R}_n(a,\epsilon,\alpha)$) 
and give a relation between $\epsilon$ and $\alpha$,
which we write $B_n(\epsilon,\alpha)=0$
(respectively $\hat{B}_n(\epsilon,\alpha)=0$).
The curves $B_n(\epsilon,\alpha)=0$ are shown in
fig.(\ref{fig rychkov}) for several values of $n$. The
curve $B(\epsilon,\alpha)=0$, calculated from
numerical integration of the system, is also
given. In figure (\ref{fig rychkov2}), we show the 
curves $\hat{B}_n(\epsilon,\alpha)=0$ and $B(\epsilon,\alpha)$.
Again, the improvement obtained with the
polynomials $\hat{R}_n(a,\epsilon,\alpha)$
is very important. The curves 
$\hat{B}_n(\epsilon,\alpha)=0$ represent
better approximations to the curve 
$B(\epsilon,\alpha)=0$ than the curves
$B_n(\epsilon,\alpha)=0$ do.

It can be proved by perturbation methods
that the asymptotic behavior of the function
$B(\epsilon,\alpha)$ when $\epsilon \to \infty$
is given by $B(\epsilon,\alpha) \sim \alpha -5$.
The curves $B_n(\epsilon,\alpha)=0$ have not
this asymptotic behavior when $\epsilon \to \infty$.
On the contrary, the curves $\hat{B}_n(\epsilon,\alpha)=0$
that we have studied ($n$ between 2 and 20) have a
correct asymptotic behavior (see fig.(\ref{fig rychkov2})).
In this way, the curves $\hat{B}_n(\epsilon,\alpha)=0$
have the right global shape when compared to the numerical
bifurcation curve. Both families of curves 
$B_n(\epsilon,\alpha)=0$ and $\hat{B}_n(\epsilon,\alpha)=0$
give lower bounds to the unknown exact bifurcation
curve $B(\epsilon,\alpha)=0$. For the curves
 $B_n(\epsilon,\alpha)=0$, this result has been established
in \cite{h et s2}. But for the curves
 $\hat{B}_n(\epsilon,\alpha)=0$ it is an 
{\it experimental} fact that cannot be
proved in a simple way.\\
Let us point out that, despite the fact that the curves
$B_n(\epsilon,\alpha)=0$ have not the correct
asymptotic behavior
, for a given value of $\alpha$, the value of $\epsilon$
obtained from the equation $B_n(\epsilon,\alpha)=0$ can be as
near as we want to the exact value of the bifurcation curve
provided that we take $n$ sufficiently large.\\

In summary, the curves $\hat{B}_n(\epsilon,\alpha)=0$ 
represent a
sequence of algebraic approximations to the bifurcation
curve $B(\epsilon,\alpha)=0$. These approximations are
 very good,
even for small values of $n$. They are better than
the approximations given by the exact lower bounds
$B_n(\epsilon,\alpha)=0$.
The improvement obtained from the polynomials
$\hat{R}_n(a,\epsilon,\alpha)$ is very surprising
because it seems that it is a general fact, valid
for arbitrary odd polynomials $F(x)$. The
mathematical justification of this method
(more specifically the passage from the polynomials
$R_n(x,\epsilon,\alpha)$ to the ``averaged'' polynomials
$\hat{R}_n(a,\epsilon,\alpha)$) represent an interesting
open problem.

In the meantime, the method presented in this paper
gives a very effective way for obtaining information
about the number of limit cycles, their amplitudes
and their bifurcations for the Li\'enard systems.

\clearpage

\begin{table}[t]
\begin{center}
\begin{tabular}{|c||c|c|c|c|c|c|c|c|c|c||c|}
\hline
n & 2 & 4 & 6 & 8 & 10 & 12 & 14 & 16 & 18 & 20 & num.\\
\hline
root & 1.732 & 1.819 & 1.863 & 1.890 & 1.909 & 1.923 & 1.934
 & 1.943 & 1.950 & 1.955 & 2.023\\
\hline
\end{tabular}
\end{center}
\caption{Values of the roots of $R_n(x,\epsilon)$ for system
(\ref{eq vdp}) with $\epsilon=3$}
\label{tbl vdp1}
\end{table}

\begin{table}[t]
\begin{center}
\begin{tabular}{|c||c|c|c|c|c|c|c|c|c|c||c|}
\hline
n & 2 & 4 & 6 & 8 & 10 & 12 & 14 & 16 & 18 & 20 & num.\\
\hline
root & 2 & 2 & 2.003 & 2.006 & 2.008 & 2.010 & 2.011
 & 2.012 & 2.013 & 2.014 & 2.023 \\
\hline
\end{tabular}
\end{center}
\caption{Values of the roots of $\hat{R}_n(a,\epsilon)$ for system
(\ref{eq vdp}) with $\epsilon=3$}
\label{tbl vdp2}
\end{table}

\end{document}